\documentclass[ %
aip,
 amsmath,amssymb,
 reprint,%
]{revtex4-1}

\usepackage[utf8]{inputenc}
\usepackage[T1]{fontenc}
\usepackage{mathptmx}
\usepackage{amsmath}
\usepackage{amsfonts}
\usepackage{amssymb,amscd,amsbsy}
\usepackage{epsfig}
\usepackage{color}
\usepackage{graphicx}
\usepackage{bm}
\usepackage{subcaption}
\usepackage{float}

\usepackage{mathtools}
\usepackage{epstopdf}
\DeclareGraphicsExtensions{.ps}
\graphicspath{{./}}

\newcommand{\calF}{ {\cal F} }
\newcommand{\ex}{ {\rm ex} }
\newcommand{\ext}{ {\rm ext} }
\newcommand{\id}{ {\rm id} }
\newcommand{\rhoeq}{ \rho^{\mbox{\tiny eq}} }
\newcommand{\rhoml}{ \rho^{\mbox{\tiny ML}} }

\newcommand{\rhomc}{ \rho^{\mbox{\tiny MC}} }

\newcommand{\FML}{ {\cal F}^{\mbox{\tiny ex,ML}} }
\newcommand{\FMLatt}{ {\cal F}_\text{\tiny att}^{\mbox{\tiny ex,ML}} }

\newcommand{\FHR}{ {\cal F}^{\mbox{\tiny HR}} }

\newcommand{\Fatt}{ {\cal F}^\text{\tiny ex}_{\mbox{\tiny att}} }
\newcommand{\fatt}{ {\mathfrak{f}}^\text{\tiny ex}_{\mbox{\tiny att}} }
\newcommand{\Fex}{ {\cal F}^{\mbox{\tiny ex}} }
\newcommand{\fML}{ {\mathfrak{f}}^{\mbox{\tiny ex,ML}} }

\newcommand{\mueq}{ {\mu}^{\mbox{\tiny eq}} }
\newcommand{\muML}{ {\mu}^{\mbox{\tiny ML}} }

\newcommand{\dcf}{ C^{(2)} }

\newcommand{\nw}{n_\text{w}}

\newcommand{\bigv}{\biggr\rvert}

\DeclareSymbolFont{newfont}{OML}{cmm}{m}{it}
\DeclareMathSymbol{\Epsilon}{3}{newfont}{15}

\begin{document}
\title{Analytical classical density functionals from an equation learning network}

\author{S.-C. Lin}
 \email{shang-chun.lin@uni-tuebingen.de}
\affiliation{Institut f\"ur Angewandte Physik, Eberhard Karls Universit\"at T\"ubingen, 72076 T\"ubingen, Germany}
\author{G. Martius}
\affiliation{Max Planck Institute for Intelligent Systems T\"ubingen,
  72076 T\"ubingen, Germany}
\author{M. Oettel}
\affiliation{Institut f\"ur Angewandte Physik, Eberhard Karls Universit\"at T\"ubingen, 72076 T\"ubingen, Germany}

\begin{abstract}
We explore the feasibility of using machine learning methods to obtain an analytic form of the
classical free energy functional for two model
fluids, hard rods and Lennard--Jones, in one dimension.
The Equation Learning Network proposed in Ref.~\onlinecite{Martius2016} is suitably modified to construct
free energy densities which are functions of a set of weighted densities and which are built from a small number of basis
functions with flexible combination rules. This setup considerably enlarges the functional space
used in the machine learning optimization
as compared to previous work \cite{Lin2019} where the functional is limited to a simple polynomial form.
As a result, we find a good approximation for the exact hard rod functional and its direct correlation
function. For the Lennard--Jones fluid, we let the network learn (i) the full excess free energy functional
and (ii) the excess free energy functional related to interparticle attractions.
Both functionals show a good agreement with simulated density profiles for thermodynamic parameters inside and
outside the training region.
\end{abstract}
\maketitle

\section{Introduction}
\label{sec:intro}

Density functional theory (DFT) may be viewed as a great reductionist scheme for classical and quantum many--body systems in equilibrium.
The one--to--one correspondence between the one--body density profile of particles and the one--body external
potential acting on these particles entails that a unique (free) energy functional of the one--body density
contains all of the homogeneous and inhomogeneous equilibrium structure in a given system, and no explicit knowledge
of higher--order correlations (i.e. through the phase space distribution of classical particles or the full
many--body quantum wavefunction) is needed.

In general, the analytical form of the (free) energy functional is unknown, except for a handful of particular
model systems (mostly in one dimension [1D]). In recent years, some effort has gone into approximating (``learning'') functionals
by machine learning (ML) techniques. In quantum DFT, e.g., interpolating functionals generated by kernel ridge regression
have been tested for model 1D systems \cite{Burke2012,Burke2016} and also have been extended to 3D systems \cite{Burke2017}. Numerically interpolated
functionals do not contain sufficient information about functional gradients, therefore both the
energy--density map and the external potential--density maps had to be learned by interpolation \cite{Burke2016}. For the 1D Hubbard model,
a convolutional network functional has been learned whose numerical functional derivative appears to be more robust~\cite{Sanvito2019}.
However, these approaches hide the energy functional inside an ``ML black box'' which does not permit much insight from
a theory perspective. For the classical case, a 1D LJ like fluid was studied with a convolutional network~\cite{Lin2019} ,
utilizing an established approach from liquid state theory of splitting the excess free energy functional into a ``repulsion'' part
 and an ``attraction'' part $\Fatt$ \cite{Lin2019}. The convolutional network naturally
led to an approximation of $\Fatt$ in terms of weighted densities $n_i$, which are the essential building blocks in modern classical DFT;
however, the free energy density $\fatt(n_i)$ as a function of  $n_i$ had to be prescribed as simple polynomials.
An interpretable results obtained in \cite{Lin2019} was the accurate splitting of the interaction potential in the Weeks--Chandler--Andersen (WCA) spirit \cite{Bishop1984}.

In this context, the question naturally arises whether ML techniques can be used to learn analytic forms of (free) energy functionals
instead of ``black boxes'' or presumed forms. This question is important also in a more general context: can ML algorithms contribute
to theory building in physics? In the ML community, efforts in that direction have utilized genetic algorithms to search a
space of simple basis function with multiplication and addition rules~\cite{SchmidtLipson2009}.
More recent work proposes an equation learning network employing gradient-based optimization with simple basis functions and division besides multiplication/addition as combination rules~\cite{Martius2016,Martius2018}.
An empirical  principle for the ``right'' formula (choose the simplest one that still predicts well, i.e. Occam's razor) can be built into the cost function.
This principle was also successful in the history of physics in finding analytical models with high predictive power even outside the training/observed 
regime.
For the DFT problem, the extrapolation power to other external potentials
is an important aspect, as well as the analytic differentiability of the free energy functional since
structural information about the fluid (pair correlations)
is obtained via the direct correlation function (two functional derivatives of the excess free energy functional).
These aspects are explored below for the model cases of a hard rod (HR) and a Lennard--Jones (LJ) fluid in 1D.

\begin{center}
\begin{figure*}[t]
\includegraphics[width=0.8\textwidth]{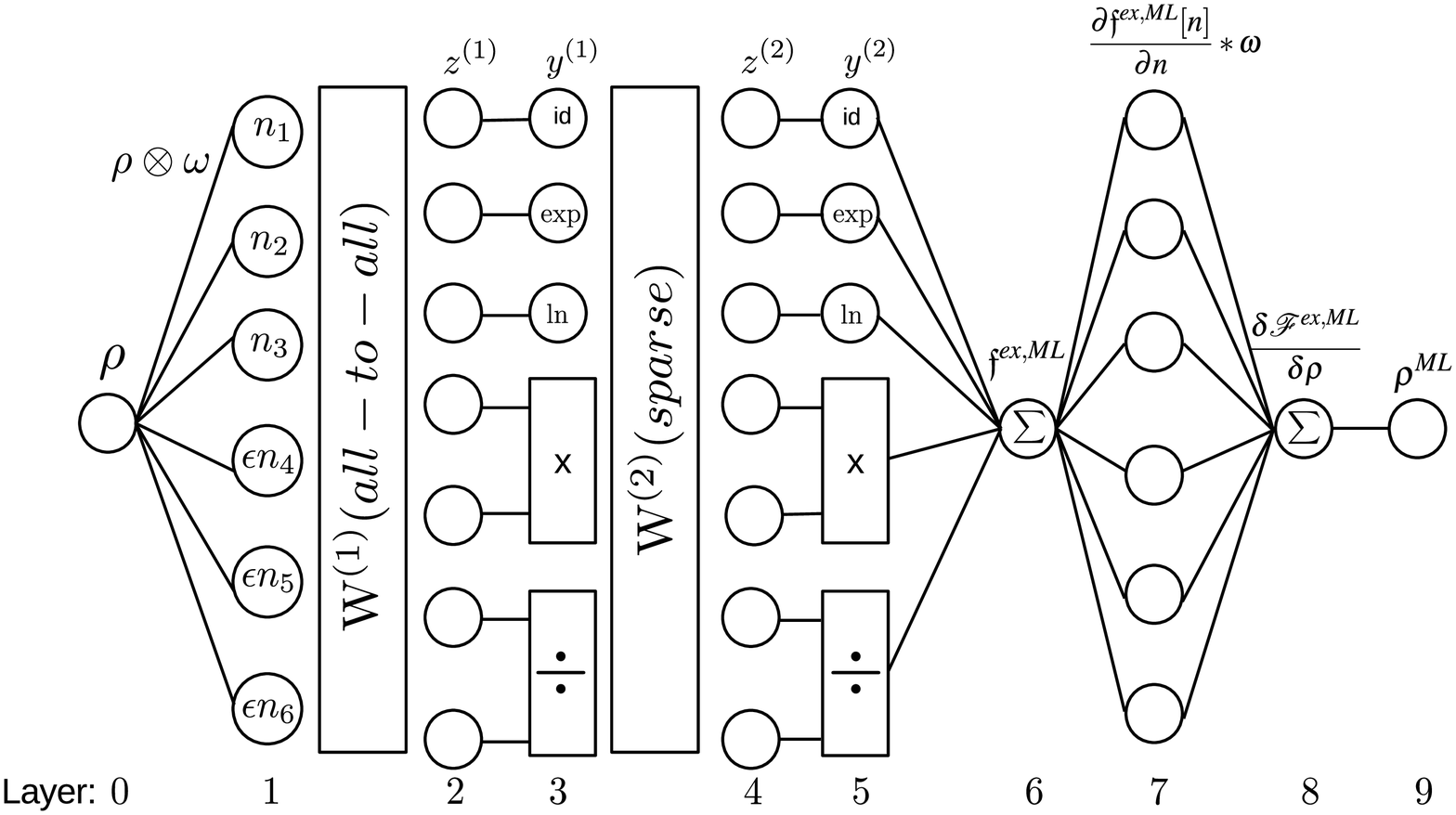}
\caption{Network architecture of the proposed FEQL for 10 layers ($L=9$) and one neuron per type ($u = 3$, $v = 2$) and 6 convolution (weighting) kernels ($\nw=6$).
$\Epsilon$ is the coupling strength (equivalent to inverse temperature) in the LJ potential.
}
\label{fig:model}
\end{figure*}
\end{center}


\section{Classical DFT}
\label{sec:DFT}
In classical DFT \cite{hansen2013theory,evans1979nature,henderson1992fundamentals,Roth2010}, the grand potential functional of 1D system is
\begin{equation}
\label{eqn:cDFT_101}
\Omega[\rho(x)] = \calF^{\id}[\rho(x)]+\calF^{\ex}[\rho(x)]+\int d x ({V}^{\ext}(x)-\mu) \rho(x),
\end{equation}
where $\rho(x)$ is the particle density distribution,  $\calF^{\id}$ is the free energy functional of the ideal gas,
$\calF^{\ex}$ is the excess free energy functional (unique for a given pair potential between particles), $\mu$ is the chemical potential and
$V^{\ext}$ is an external potential.  The exact form of $\cal F^{\id}$ is:
\begin{equation}
\label{eqn:free_idea_gas}
\beta \calF^{\id}=\int\,d x\,\rho\left(x\right)\left[ \ln\left(\rho\left(x\right)\lambda\right)-1\right]
\end{equation}
with $\beta=1/(k_\text{B} T)$, $T$ the temperature, $k_\text{B}$ Boltzmann's constant, and $\lambda$ the thermal wavelength. In the following we set $\beta=\lambda=1$.

In equilibrium,  the corresponding density profile $\rhoeq$ minimizes $\Omega$ for a given $\mu$. Thus, with $\frac{\delta\Omega}{\delta \rho}=0$ and
Eq.~\eqref{eqn:free_idea_gas}, we obtain
\begin{equation}
\rhoeq= \exp\left(\mu-\frac{\delta \calF^{\ex}}{\delta \rho}\bigv_{\rho=\rhoeq}-V^{\ext}\right).
\label{eqn:rho_eq}
\end{equation}
All DFT solutions for test density distributions in this work are obtained by iteratively solving Eq.~\eqref{eqn:rho_eq}
using the Picard method with mixing.

In this paper, we investigate the HR pair potential:
\[
    U_{\rm{HR}}(x)=
\begin{cases}
	\infty &\text{if}\hspace{2em} x<\sigma\\
    0 &\text{otherwise}
\end{cases}
\]
as well as  the LJ(--like) potential:
\[
    U_{\rm{LJ}}(x)=
\begin{cases}
	\infty &\text{if}\hspace{2em} x<\sigma\\
    4\Epsilon\left[\left(\frac{\sigma}{x}\right)^{12}-\left(\frac{\sigma}{x}\right)^{6}\right]  &\text{if}\hspace{2em} \sigma<x<16\sigma\\
    0 &\text{otherwise}
\end{cases}
\]
with $x$ the distance between particle centers, $\sigma$ the diameter of the particles and $\Epsilon$ the strength of interaction. In the following we set $\sigma=1$.

%
%
%
%

\section{Machine learning}
\label{sec:model}
\subsection{Model}

\begin{figure*}[t]
\begin{subfigure}[c]{0.32\textwidth}
\includegraphics[width=\textwidth]{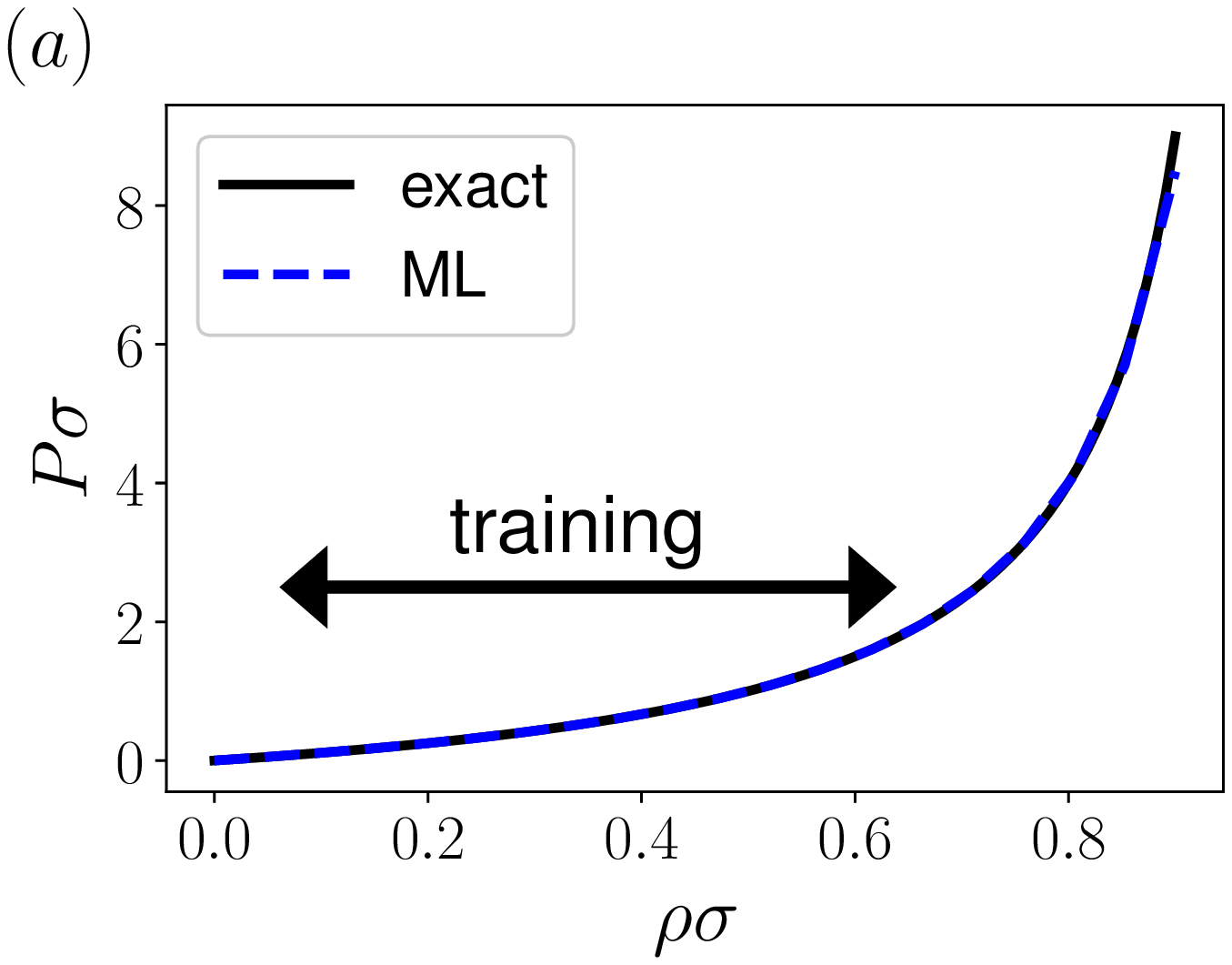}
\end{subfigure}
\begin{subfigure}[c]{0.34\textwidth}
\includegraphics[width=\textwidth]{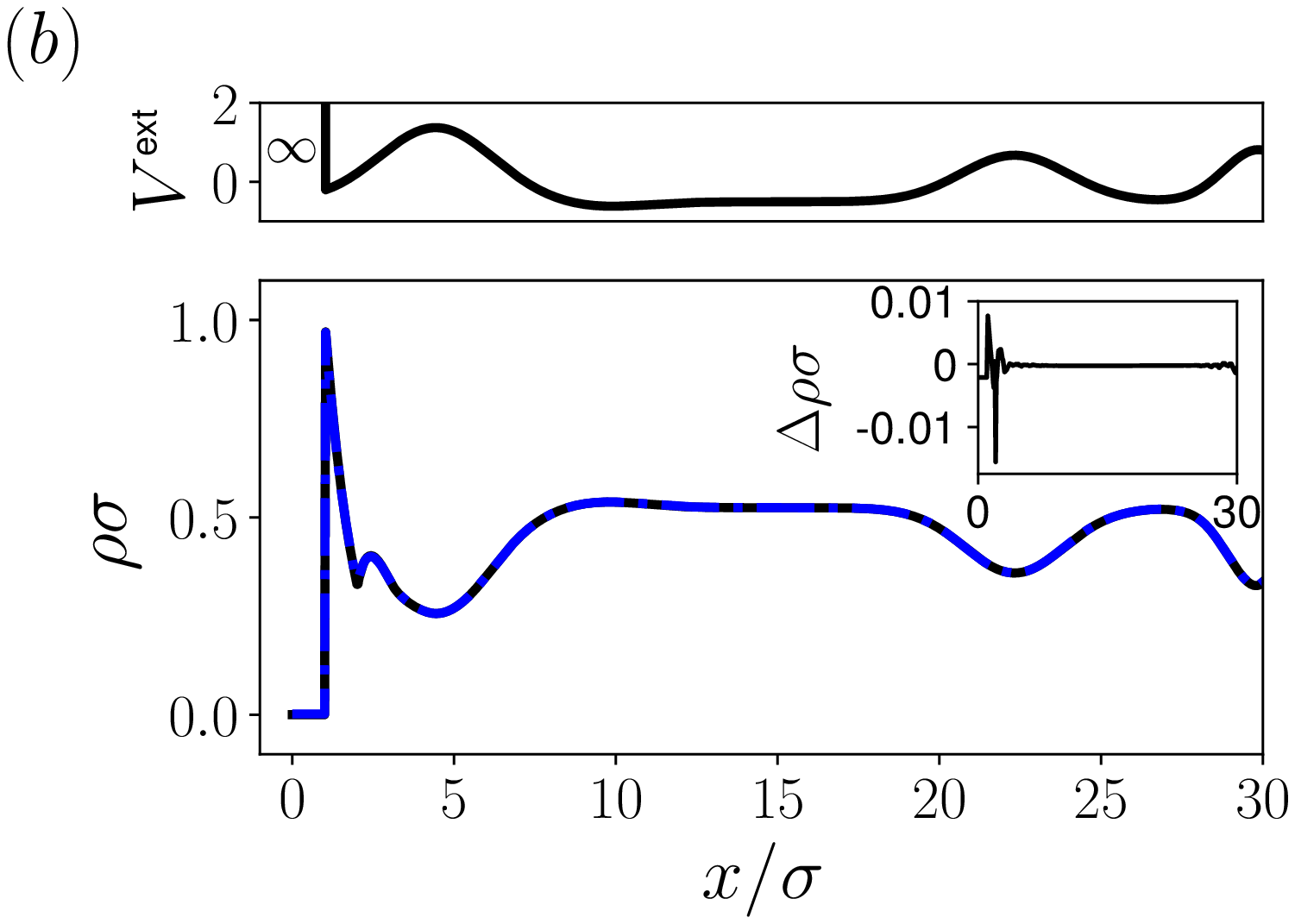}
\end{subfigure}
\begin{subfigure}[c]{0.32\textwidth}
\includegraphics[width=\textwidth]{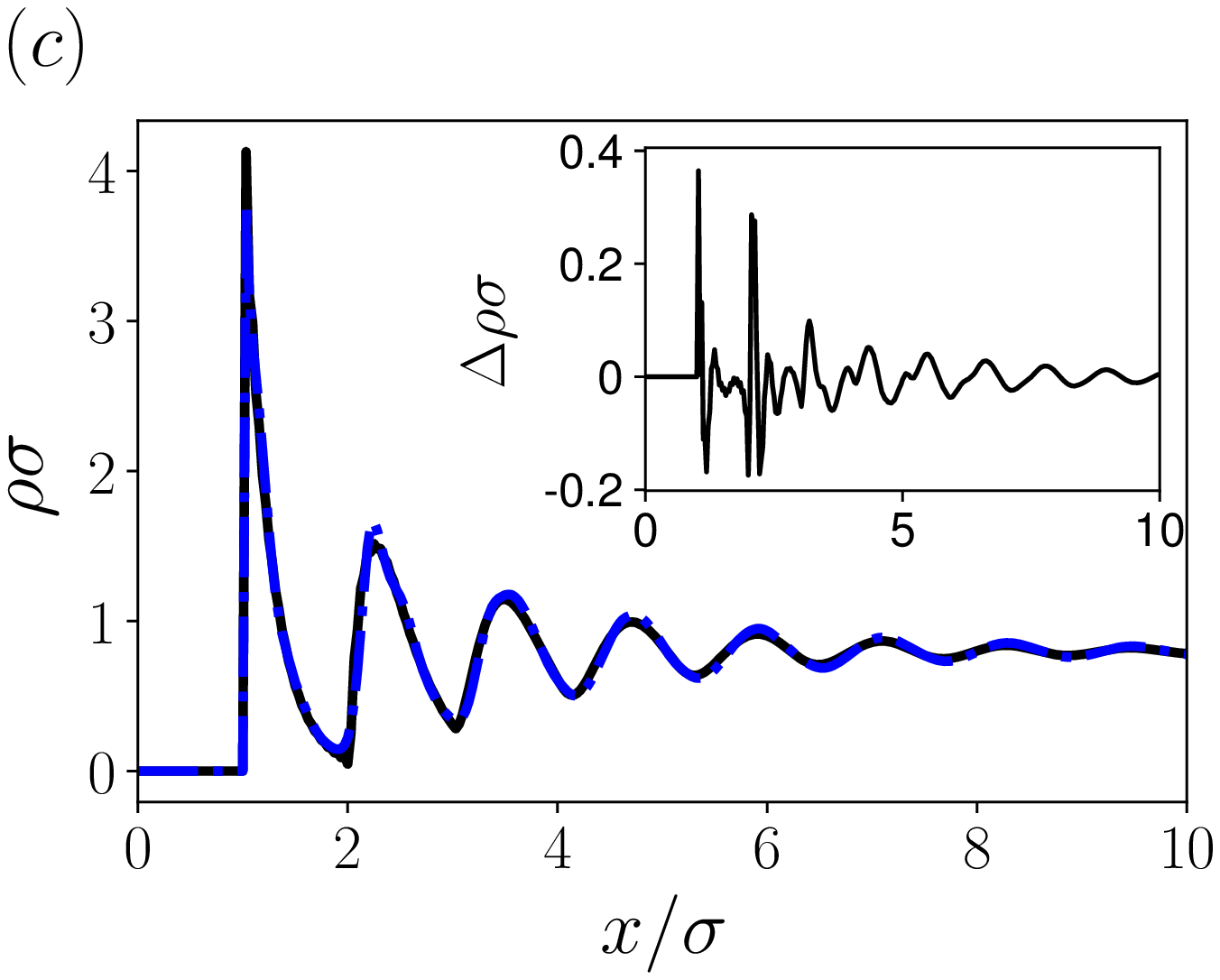}
\end{subfigure}
\caption{FEQL results for hard rods. Dark solid lines are exact solutions from $\FHR$ and blue dashed lines are ML results. (a) eos, $P(\rho)$.
(b) density profile for $\rho_0=0.49$ inside the training region but $V^{\ext}$ not in the training data.
(c) density profile at hard wall for $\rho_0=0.80$ outside the training region. Insets in (b) and (c) show $\Delta\rho=\rho^{\rm exact}-\rhoml$.
}
\label{fig:HR_result}
\end{figure*}

We define a machine learned excess free energy functional $\FML$ and
the resulting ML output density $\rhoml$ by
\begin{equation}
\label{eqn:rho_ML}
\rhoml(x) = \exp\left(\muML-\frac{\delta \FML}{\delta \rho}\bigv_{\rho=\rhoeq}-V^{\ext}\right).
\end{equation}
This is the equivalent of a ``generative step'' of a learned distribution $\rhoml$ from an input distribution
$\rhoeq$ in an ML network (via weighted densities $n_i$, see below).
Here, $\muML$ is included in the training process to facilitate convergence. In the end, $\muML \to \mu$, see also
 Ref.~\onlinecite{Lin2019} and SI for details.
 The test density profiles shown in Figs.~\ref{fig:HR_result}--\ref{fig:LJ_result} are obtained 
 by initializing with a constant value and then iteratively minimizing the learned functional with the (physical) $\mu$. 

The network we propose, Functional Equation Learner (FEQL), is a $L$--layered feed--forward network with computational units specifically designed for constructing
the free energy functional (see Fig.~\ref{fig:model}).
The first layer consists of convolution kernels which compute the weighted densities
$n_i$ with the convolution kernel $\omega_i$ ($i=1...\nw$) by
\begin{equation}
n_i(x)=\rho\otimes \omega_i = \int \,dx^{\prime}\, \rho(x^{\prime})\omega_i(x-x^\prime)\;,
\end{equation}
and some of the weighted densities are multiplied by $\Epsilon$ in the case of the LJ fluid.
The maximum allowed ranges of the $\omega_i$  are $[-4\sigma,4\sigma]$ (HR) and $[-8\sigma,8\sigma]$ (LJ).
Using weighted densities instead of the particle density is inspired by the exact HR functional \cite{percus1976} and
fundamental measure theory \cite{rosenfeld1989free, Roth2010}. The layer 2 is a linear, all--to--all mapping of the vector
(of functions) $n=\{n_i(x)\}$ to the vector 
\begin{equation}
z^{(l=1)}=W^{(1)} n
\end{equation}
at level $l=1$. The layers $3...L-4$ are a sequence of nonlinear and linear transformations.
The non-linear transformation at level $l$
contains $u$ unary units $f_I$ and $v$ binary units $g_J$ and maps $z^{(l)}(x)$ ($u+2v$--dimensional) to the layer output $y^{(l)}$ ($u+v$--dimensional) as:
\begin{eqnarray}
y^{(l)}:=&&\left(f_1\left(z_1^{(l)}\right),f_2\left(z_2^{(l)}\right)...f_u\left(z_u^{(l)}\right),\right.\nonumber\\
&&\left(g_1\left(z_{u+1}^{(l)},z_{u+2}^{(l)}\right)...g_v\left(z_{u+2v-1}^{(l)},z_{u+2v}^{(l)}\right)\right).
\end{eqnarray}
The unary units, $f_1, . . . , f_u$ receive the respective component, $z_1, . . . , z_u$ as inputs, and each unit is one of the following base functions indexed by $I \in {0, 1, 2}$:
\begin{align*}
f_I(z_i):=\begin{cases}
z_i  \quad        &\mbox{if} \quad I=0 \\
\exp(z_i)-1  \quad  &\mbox{if} \quad  I=1 \\
\ln(z_i+1)  \quad   &\mbox{if} \quad  I=2 \\
\end{cases}
\end{align*}
The binary units, $g_1, . . . , g_v$ receive the remaining component, $z_{u+1}, . . . , z_{u+2v}$, as input in pairs of two,
and each unit may be multiplication or division indexed by $J \in {0, 1}$:
\begin{align*}
g_J(z_i,z_{i+1}):=\begin{cases}
z_i\times z_{i+1}  \quad  &\mbox{if} \quad  J=0 \\
z_i\div \left(z_{i+1}+1\right)  \quad   &\mbox{if} \quad  J=1 \\
\end{cases}
\end{align*}
Note that $f_I(0)=g_J(0,z)=0$. One may worry about divergences in division and logarithm when $z\rightarrow-1$. In the beginning of the training procedure, all parameters and convolution kernel are initialized by small numbers so thus $z$ is close
to zero and there are no problems. If $z$ is too close to -1, the loss will change drastically; thus the network will intrinsically handle this issue.
As mentioned in Ref.~\onlinecite{Martius2018}, one could use modified division and logarithm functions and add extra penalties. However, it turns out not to be required here. 

The linear transformation from level $l$ to $l+1$ maps the $(u+v)$--dimensional input $y^{(l)}$ to the $(u+2v)$--dimensional intermediate representation $z^{(l+1)}$ given by 
\begin{equation}
z^{(l+1)}=W^{(l+1)}y^{(l)}.
\end{equation}
Thus, the $\nw$ convolution kernels $\omega(x)$ in the first layer 
and the matrices $W^{(l)}$ are free parameters that are learned during training.

The machine-learned free energy density $\fML$ is a summation of the output of layer $L-4$, the functional derivative
 $\frac{\delta \FML}{\delta \rho}=\sum_{i}\frac{\partial \fML}{\partial n_i}\ast \omega_i$  (with $\FML=\int dx \,\fML(n)$ and $\ast$ denoting cross--correlation)
is used in the final, generative step (Eq.~\eqref{eqn:rho_ML}). More details about constructing  FEQL can be found in the SI.

\subsection{Network training}
To obtain training data for $\rhoeq$, grand canonical simulation are used in the case of LJ fluid;
for the HR fluid, Eq.~\eqref{eqn:rho_eq} is directly solved, since the exact functional is known.

FEQL is fully differentiable in its free parameters $\theta=[W,\omega]$ and can thus be trained using back--propagation.
We adopt the following loss function
\begin{eqnarray}
\label{eqn:loss}
L&=&\frac{1}{N}\sum_{i=1}^{N}\left(\alpha_1\int|\rhoeq_i-\rhoml_i|\,d\,x +\alpha_2|\mueq_i-\muML_i|\right)\nonumber\\
&&+\lambda_1 \sum_i \int\,dx\,|\omega_i|  +\lambda_2 \sum_{l,\beta\gamma}|W_{\beta\gamma}^{(l)}|,
\end{eqnarray}
with $\alpha_1=0.9$ and  $\alpha_2=0.1$. These values have been determined empirically and the exact choice is not critical. For training we choose Adam \cite{kingma2014adam} with mini-batches:
\begin{equation}
\label{eqn:Adam}
\theta_{t+1} =\theta_{t}+\mbox{Adam}\left(\frac{\partial L\left(D\left(t\right)\right)}{\partial \theta},\alpha\right)
\end{equation}
with $\alpha$ the stepsize parameter (learning rate) and $D(t)$ the data in the current mini-batch.
The choice of Adam is not critical and standard stochastic gradient descent also works. 

Following Sahoo \emph{et al.}~\cite{Martius2018}, we adopt a three--step training procedure.  At the beginning, we use no regularization
($\lambda_1=\lambda_2=0$), such that parameters can vary freely and reach reasonable starting  points. In step 2, we
switch on the regularization by setting $\lambda_1$ and $\lambda_2$ to positive finite values to sparsify the network for obtaining a simpler functional. In step 3, we clamp small parameters with $|W_{\beta\gamma}^{(l)}|<w_{\text{th}}$ to zero. In this way we keep the sparsity introduced by the lasso \cite{tibshirani1996regression} training in step 2 but make sure unbiased
parameter values are attained. In this paper we choose $\alpha=10^{-2}$ or $10^{-3}$, $\lambda_1=10^{-7}$ and $w_{\text{th}}=0.05$.

%
%
%
%
\section{Result}
\subsection{Hard rods}
The exact equation of state (eos) for the hard rod (HR) fluid is given by the pressure $P(\rho)=\frac{\rho}{1-\rho}$
and the analytic form of $\FHR$ (Percus functional) is one of the few exactly known ones \cite{percus1976, RY1997}.

The parameter of $\FML$ are trained using 1024 density profiles in a hard wall slit of width 32 $\sigma$ with 3 additional Gaussian
potentials of random strength/width and location inside the slit and with a range of training reservoir densities $\rho_0=0.2...0.55$.
We choose $\nw=3$ and (1,1,1,3,1) nodes for (identity, exponential, logarithm, multiplication and division) with $L=10$ layers
(see Fig.~\ref{fig:model}) and $\lambda_2=8\cdot 10^{-5}$ in Eq.~\eqref{eqn:loss} (results for different $\lambda_2$ and arguments for an optimal choice  are shown in the SI). $\FML$ is not of the form of the Percus functional, since the convolution kernels of the latter are Dirac delta and Heaviside step functions, which are are hard to be captured by our network. 

In Fig.\ref{fig:HR_result}, we show the eos, a density profile inside the thermodynamic training region but not in
the training data ($\rho_0=0.49$) and a density profile outside the training region ($\rho_0=0.80$). The FEQL recovers the almost exact
result inside the training region and also performs quite well even outside the training region. The ML density profiles are initialized by $\rho=0.5$ and then iteratively solved using Eq.~\eqref{eqn:rho_eq} with $\Fex=\FML$.   

The virial expansion
\begin{equation}
P^{\mbox{\tiny ML}}(\rho)\simeq \rho + 1.03 \rho^{2} + 0.71 \rho^{3}  + O\left(\rho^{4}\right)
\end{equation}
of the ML eos shows moderate deviation compared to the exact one ($\frac{\rho}{1-\rho}=\rho+\rho^2+\rho^3...$). 

 One sees that inaccuracies in these coefficients do not
 necessarily mean a poor eos. Despite inaccurate coefficients, the higher
 order terms in the learned eos
 combine appropriately to give a good representation of the exact eos.
This is understandable
 since no explicit information about virial coefficients is incorporated into
 the cost function, thus  the learning procedure has little incentive to find
 the correct coefficients.

%
%
%
%
%

\subsection{Lennard--Jones}
Here, 1115 training distributions are generated with random $\mu$ and $\Epsilon$ in the range of $0.5...1.5$ and $\ln 0.5...\ln 2$,
respectively, with $V^{\mbox{\tiny ext}}$ prescribed as in the hard rod case.
The training data are obtained by grand canonical Monte Carlo (GCMC) simulation.

\begin{figure*}[t]
\begin{subfigure}[c]{0.32\textwidth}
\includegraphics[width=\textwidth]{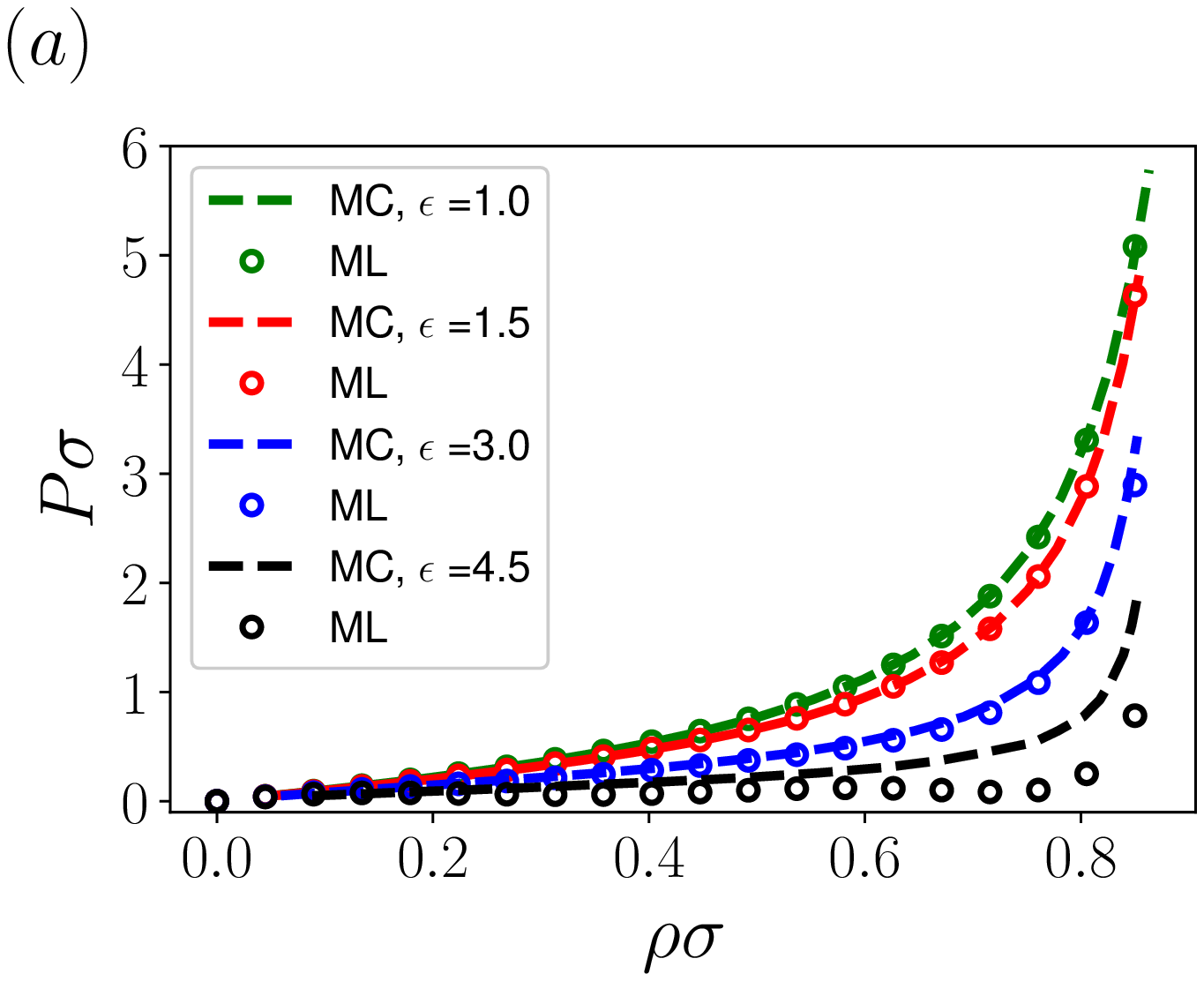}
\end{subfigure}
\begin{subfigure}[c]{0.33\textwidth}
\includegraphics[width=\textwidth]{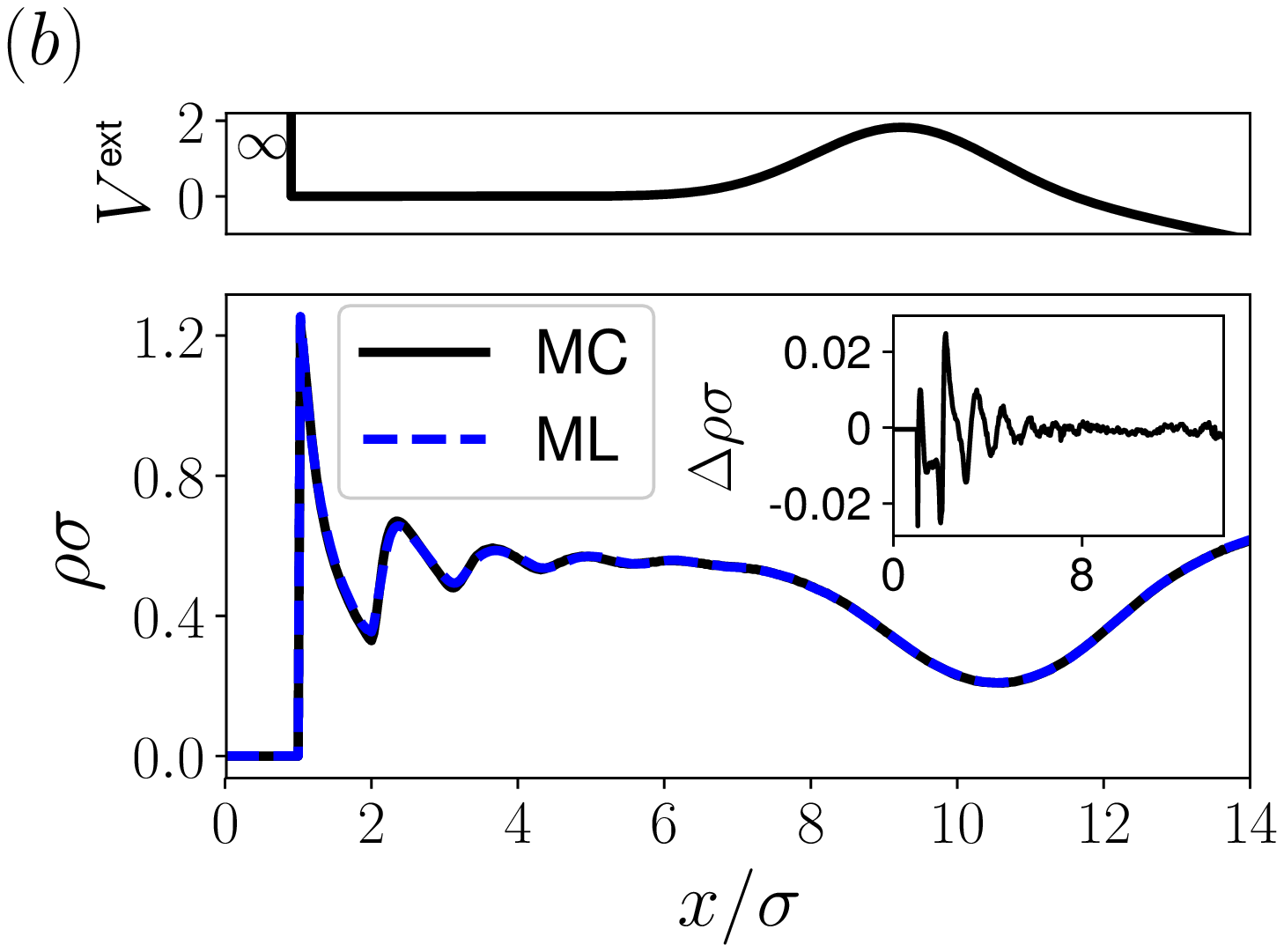}
\end{subfigure}
\begin{subfigure}[c]{0.33\textwidth}
\includegraphics[width=\textwidth]{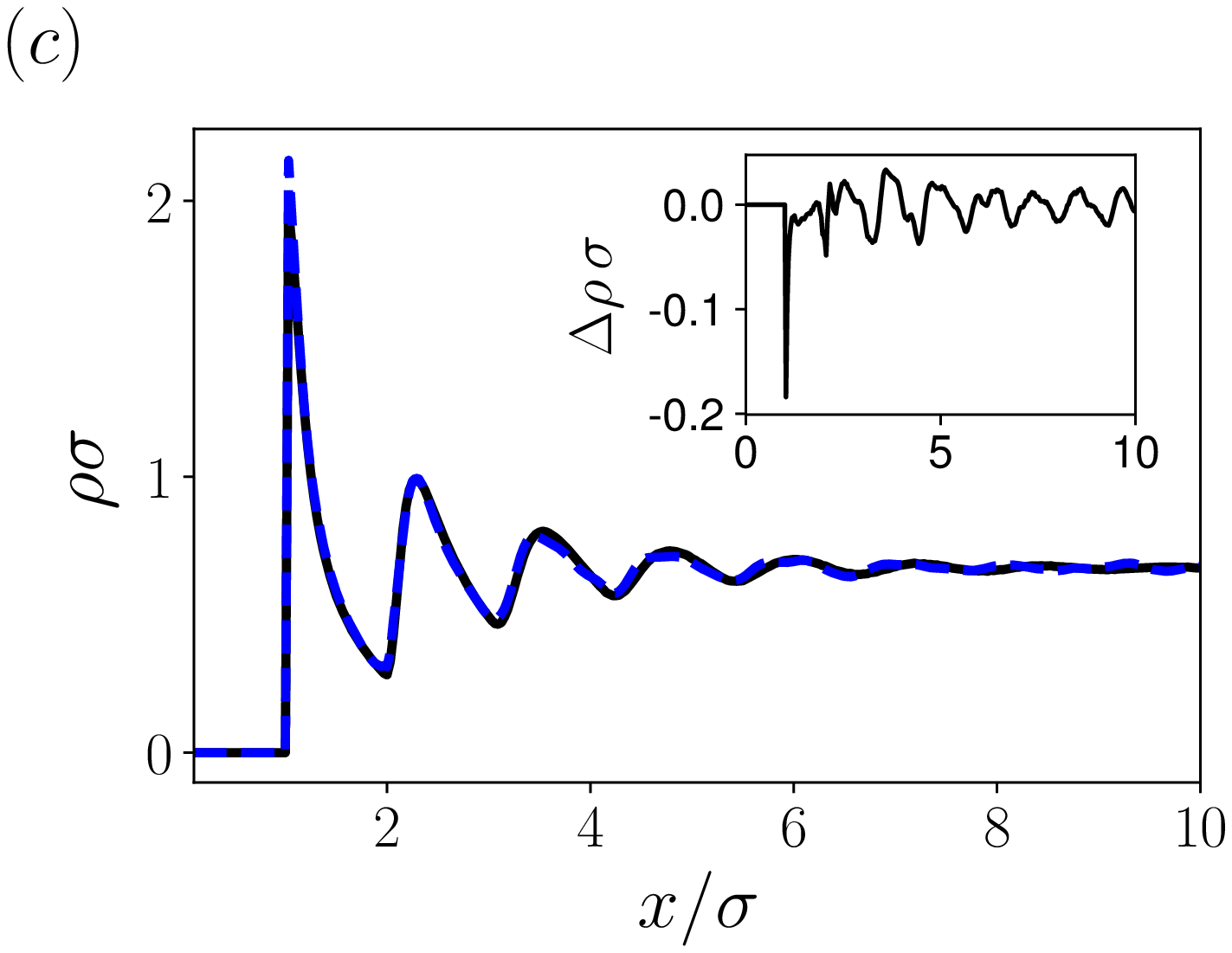}
\end{subfigure}
\caption{FEQL results for LJ fluid with functional splitting. (a) eos $P(\rho)$. (b) density profile for
$\Epsilon=1.30$, $\mu=\ln(1.27)$ inside the training region but $V^{\ext}$ not in the training data.
(c) density profile at a hard wall for $\Epsilon=1.7$, $\mu=\ln(1.7)$, outside the training region. Dark solid lines are simulation profiles
and blue dashed lines are ML results. Insets in (b) and (c) show $\Delta\rho=\rhomc-\rhoml$.}
\label{fig:LJ_tail_result}
\end{figure*}

\subsubsection{Splitting between repulsions and attractions}
Following liquid state theory \cite{Lin2019}, we split into a contribution from repulsions and one from attractions as follows:
\begin{equation}
\FML([\rho]; \Epsilon)=\FHR([\rho])+\Epsilon\FMLatt([\rho]; \Epsilon),
\end{equation}
where the factor $\Epsilon$ in front of the $\FMLatt$ makes sure $\FML(\Epsilon\to 0)=\FHR$, and $\FMLatt $ is to be learned by the network. 
In Fig.~\ref{fig:model}, the output from the layer 6 is $\FMLatt$; we multiply the output from the layer 8 by $\Epsilon$, add contribution from $\FHR$, and then feed it to the  layer 9.
In the first layer, we choose $\nw=4$, 1 kernel multiplied by $\Epsilon$ and another 3 without this factor
(see Fig.~\ref{fig:model}), and (1,1,1,2,1) nodes for (identity, exponential, logarithm, multiplication and division). 
The training parameter $\lambda_2=5 \cdot 10^{-5}$ in Eq.~\eqref{eqn:loss}. Results are shown in Fig.~\ref{fig:LJ_tail_result}.
The findings are similar to the HR case with a very good match to simulation data for the eos and test distributions
inside and outside the thermodynamic training region.  For a 1D system with hard--core repulsive and finite range attractive pair potential, the pressure must be monotonically increasing for arbitrary low temperature (high $\Epsilon$), and thus resulting no gas--liquid transition \cite{cuesta2004general}. The corresponding ML pressure shows no van der Waals (vdW) loop for attractions strengths up to $\Epsilon=4.1$, this is a qualitative step forward as compared to Ref.~\onlinecite{Lin2019}.

\begin{figure*}
\begin{subfigure}[c]{0.4\textwidth}
\includegraphics[width=\textwidth]{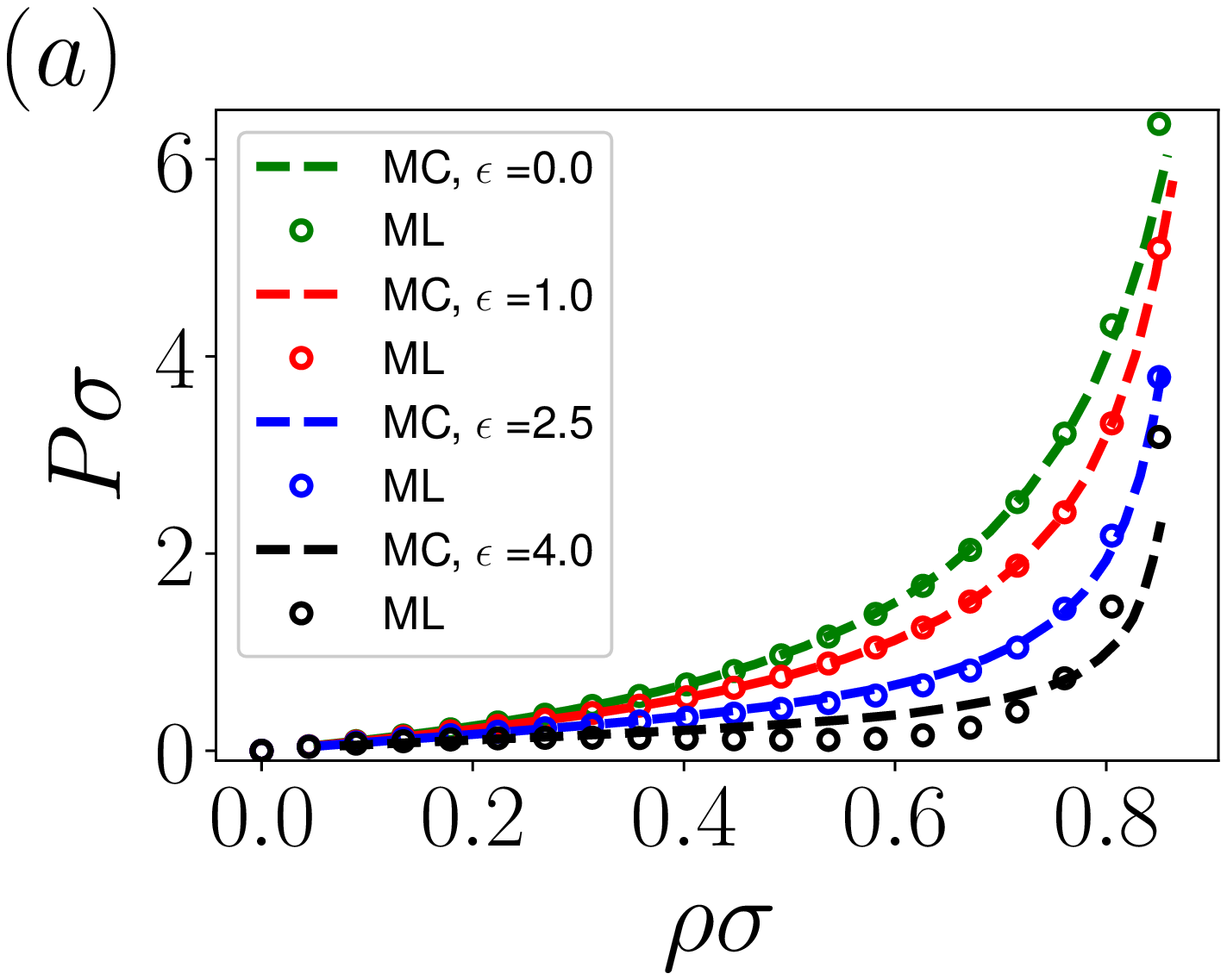}
\end{subfigure}
\begin{subfigure}[c]{0.42\textwidth}
\includegraphics[width=\textwidth]{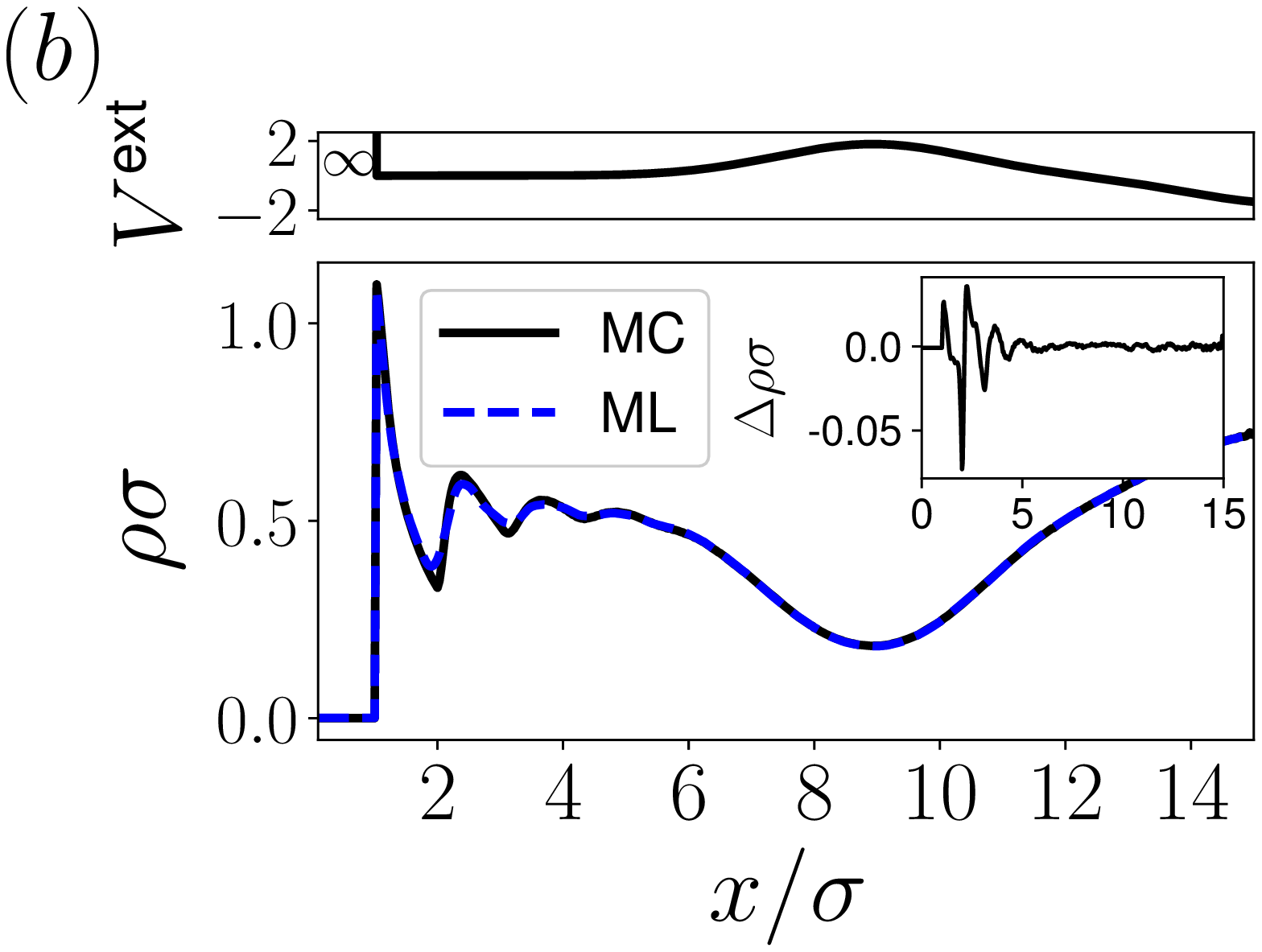}
\end{subfigure}
\begin{subfigure}[c]{0.4\textwidth}
\includegraphics[width=\textwidth]{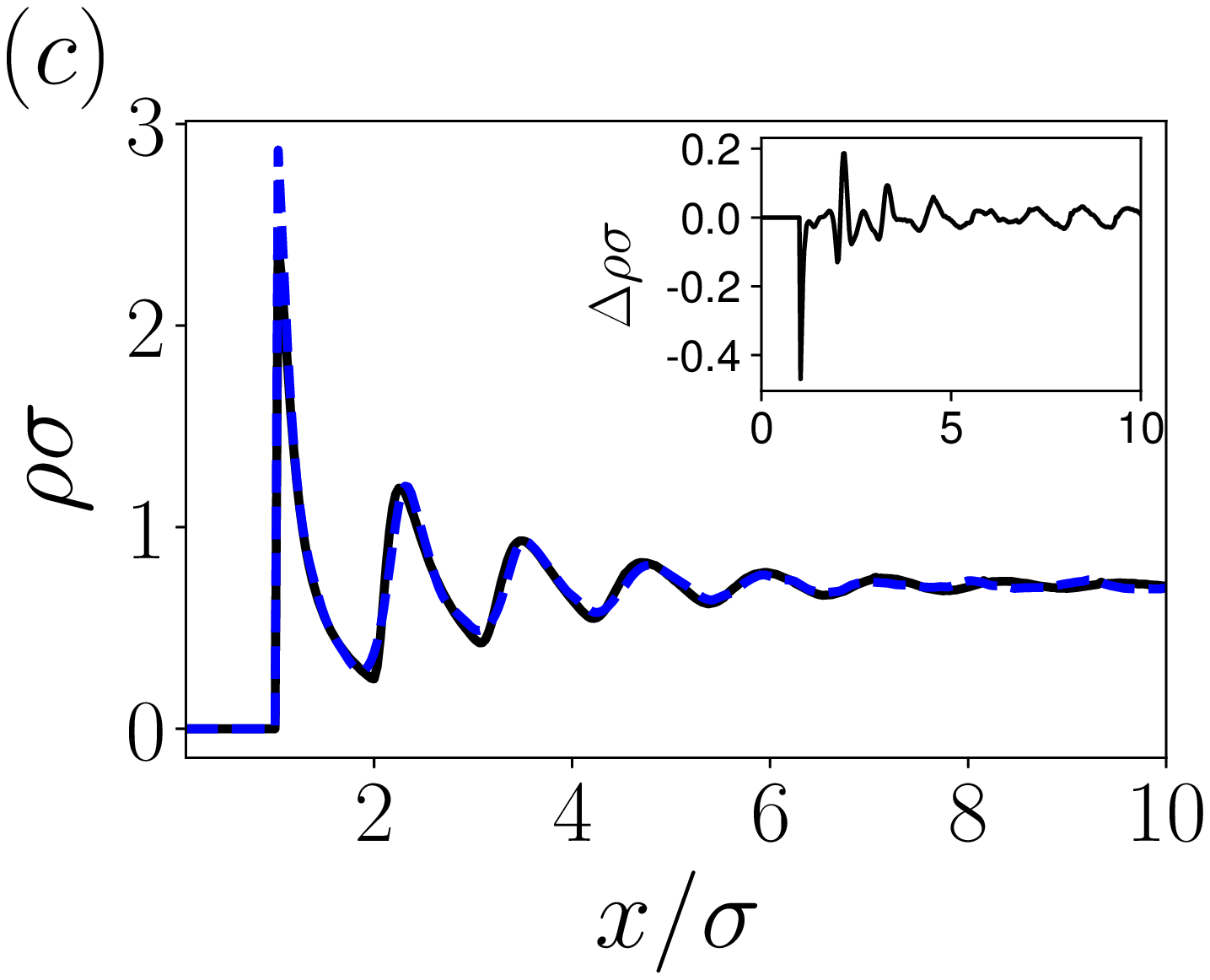}
\end{subfigure}
\begin{subfigure}[c]{0.4\textwidth}
\includegraphics[width=\textwidth]{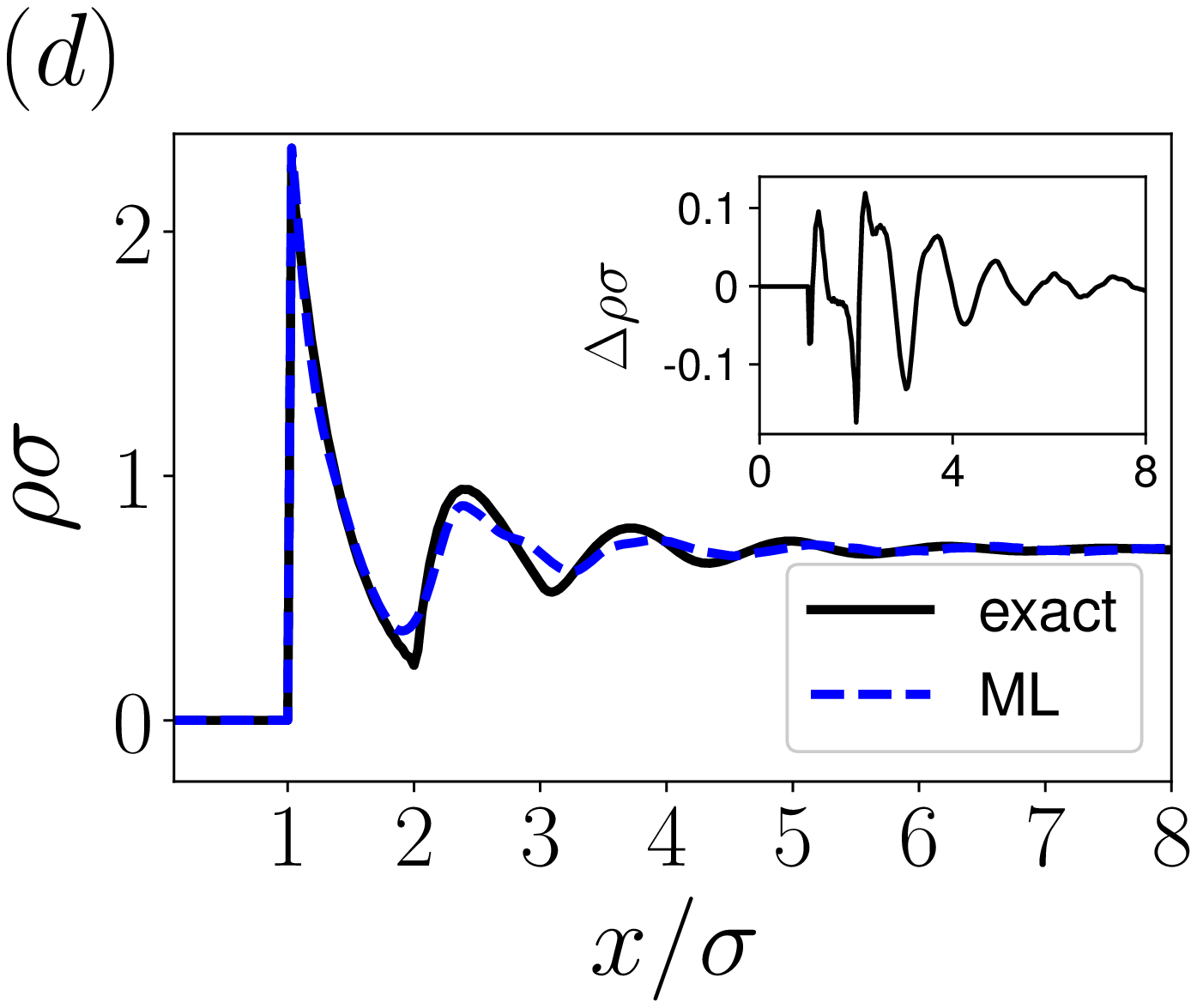}
\end{subfigure}
\caption{FEQL results for LJ fluid (no splitting). (a) eos $P(\rho)$. (b) density profile for $\Epsilon=1.25$, $\mu=\ln(1.15)$
inside the training region but $V^{\ext}$ not in the training data. (c,d) density profile at a hard wall
for $\Epsilon=1.9$, $\mu=\ln(1.9)$ (c) and $\Epsilon=0$, $\rho_0=0.7$ (HR limit, (d)).
Dark solid lines are simulation profiles and blue dashed lines are ML results.
Insets in (b) and (c) show $\Delta\rho=\rhomc-\rhoml$ and in (d) $\Delta\rho=\rho^{\rm exact}-\rhoml$.}
\label{fig:LJ_result}
\end{figure*}

\subsubsection{No splitting}
As a further test of the capability of FEQL, we forego the splitting of the functional such that
$\Fex=\FML$.
In the first layer, we choose $\nw=6$, 3 kernels multiplied with $\Epsilon$ and another 3 without this factor (see Fig.~\ref{fig:model}), and 
(1,1,1,3,1) nodes for (identity, exponential, logarithm, multiplication and division). The training parameter $\lambda_2=5 \cdot 10^{-5}$ in Eq.~\eqref{eqn:loss}.
For the training data we also include density profiles from the HR case.
In Fig.~\ref{fig:LJ_result}, we show the results. Test distributions match well to simulation data both in the HR limit and the
regime of higher attractions. The eos shows an unphysical vdW loop for attractions strengths $\Epsilon > 3.7$, much higher than the upper limit of the training data.


\subsection{Direct correlation function}
The direct correlation function (dcf) is a central object in DFT which through iterations yields the pair corelation function (Ornstein--Zernike relation, see also Chap. 3 in Ref.~\onlinecite{hansen2013theory}). It is given by the second functional derivative of $\Fex$:
\begin{equation}
\dcf(x_1,x_2;\rho_0) = -\frac{\beta \delta^2 \Fex}{\delta \rho(x_1) \delta \rho(x_2)},
\end{equation}
and it depends only on $x=|x_1-x_2|$ in the case of a homogeneous fluid with density $\rho_0$. 

\begin{figure}[t]
\centering
\includegraphics[width=0.4\textwidth]{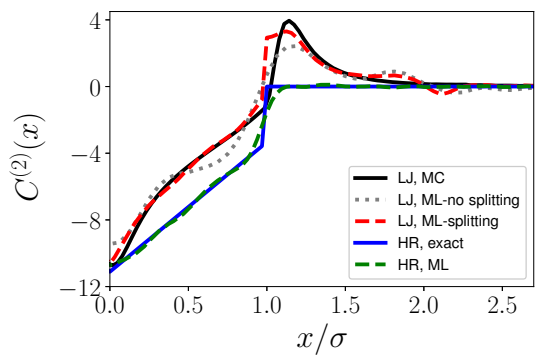}
\caption{$\dcf(x,\rho_0)$ with $\rho_0=0.7$ (HR) and 0.703 (LJ, $\Epsilon=1.85$ and $\mu=\ln(1.85)$).}
\label{fig:c2}
\end{figure}

As the network is only trained on the level of the first functional derivative (see Eq.~\ref{eqn:rho_ML}), it is a challenge for FEQL to capture the dcf.
In Fig.~\ref{fig:c2}, we show exemplary dcf's at moderate to high density for the exact HR functional, LJ from simulation and the corresponding ML results. The direct correlations inside the hard core are captured very well by ML in the HR and LJ cases. Outside the hard core, in the HR case, the $\dcf$ from the ML shows insignificant correlation. In the case of LJ, the contribution to $\dcf$ from attraction is semi--quantitatively correct, with a better result in the splitting case.    
\section{Conclusion}
The adaptation of EQL~\cite{Martius2018} to the classical DFT problem of finding $\Fex$ has shown satisfactory results
for the exemplary case of the 1D HR and LJ fluid. The new network FEQL is very flexible and goes significantly beyond
the polynomial {\em ansatz} used in Ref.~\onlinecite{Lin2019}. The analytic form allows for more easily transferable output and
further calculations to obtain, e.g., direct correlation functions.
An application to more realistic systems in 3D and perhaps also complex fluids such as water appears to be promising \cite{sergiievskyi2014fast,jeanmairet2013molecular,jeanmairet2013molecular2}. From the results of this work we conclude that the incorporation
of results from liquid state theory (separation of repulsion and attraction)
is not essential here; however, it increases the reliability and trainability of the ML functional.
Future work should include information
on virial or high density expansions as well as correlation functions (via test particles) and should develop more quantitative measures for extrapolative capabilities of
ML functionals.
\\
\section*{Supplementary Information}
See supplementary information for more discussion about the training procedure, FEQL, $\muML$, comparison of Ref.~\onlinecite{Lin2019}, explicit $\FML$ and convolution kernels $\omega(x)$.

\begin{acknowledgments}
The authors acknowledge support by the High Performance and Cloud Computing Group at the Zentrum f\"ur Datenverarbeitung of the University of T\"ubingen, the state of Baden-W\"urttemberg through bwHPC and the German Research Foundation (Deutsche Forschungsgemeinschaft, DFG) through grant no INST 37/935-1 FUGG. Further, 
We acknowledge the financial support from DFG through the fund ZUK 63 and grant OE 285/5-1 .
\end{acknowledgments}

\bibliographystyle{apsrev4-1}
\bibliography{dft_ml}

\pagenumbering{gobble}
\begin{figure}
\includegraphics[page=1,width=\textwidth]{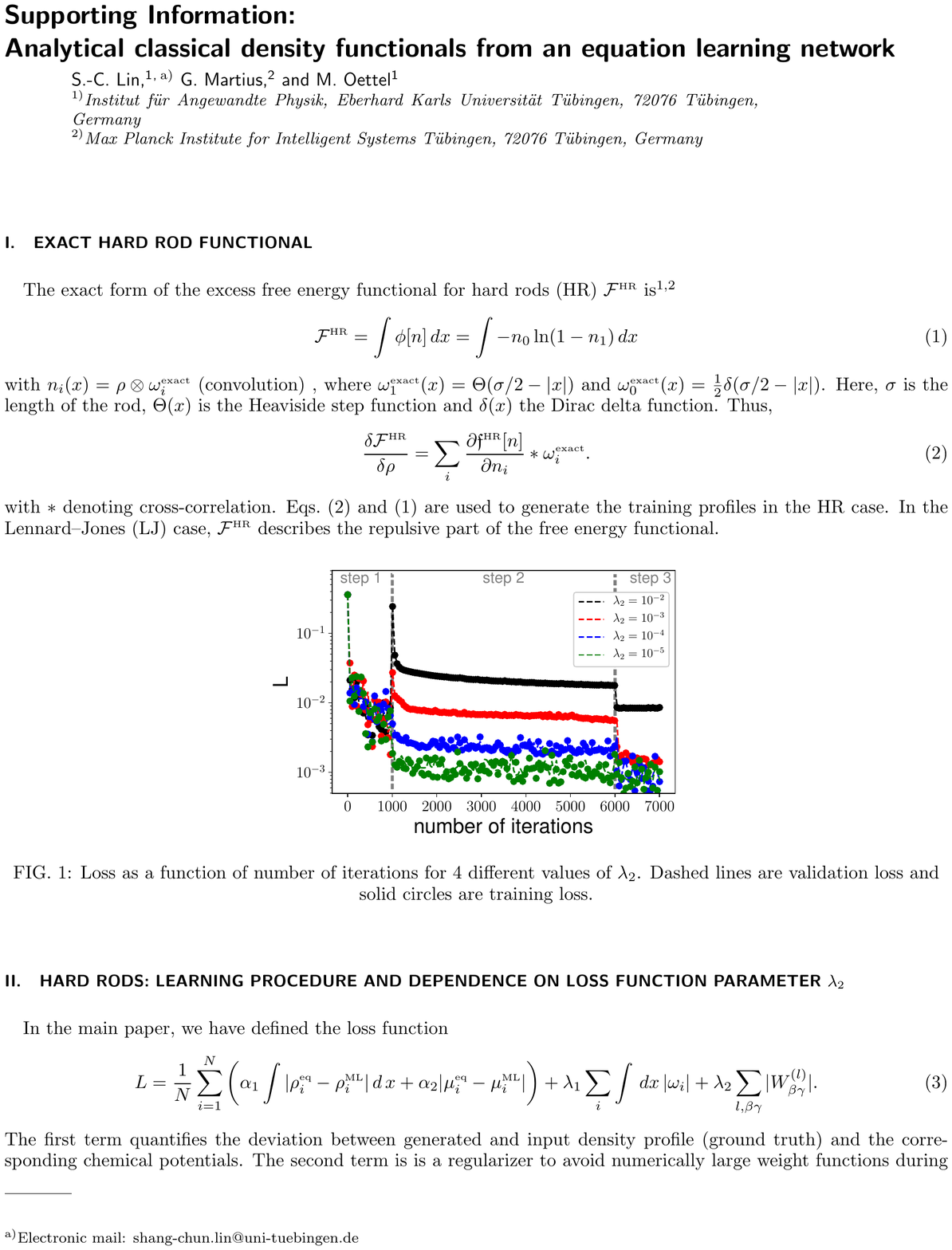}
\end{figure}
\begin{figure}
\includegraphics[page=2,width=\textwidth]{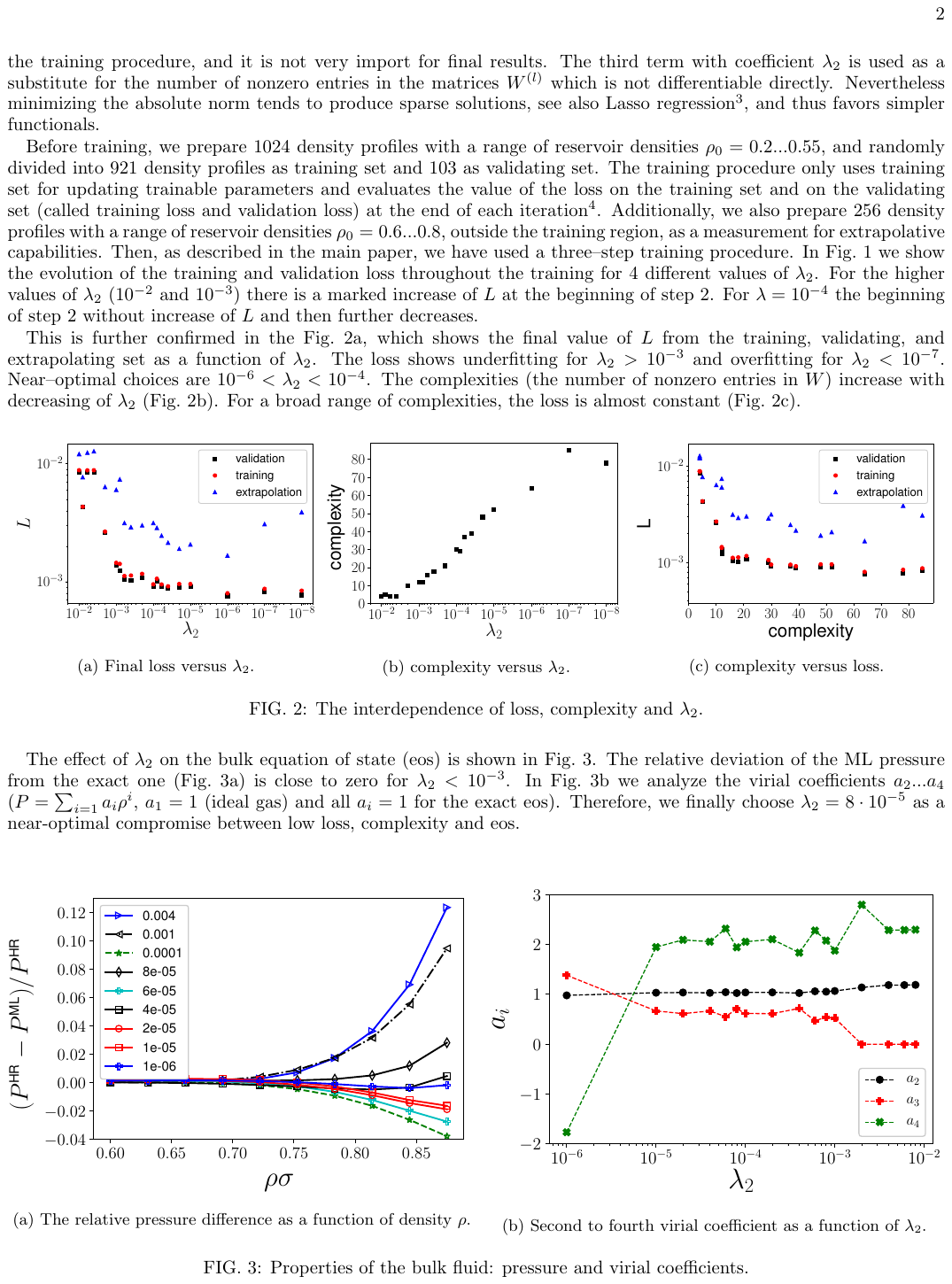}
\end{figure}
\begin{figure}
\includegraphics[page=3,width=\textwidth]{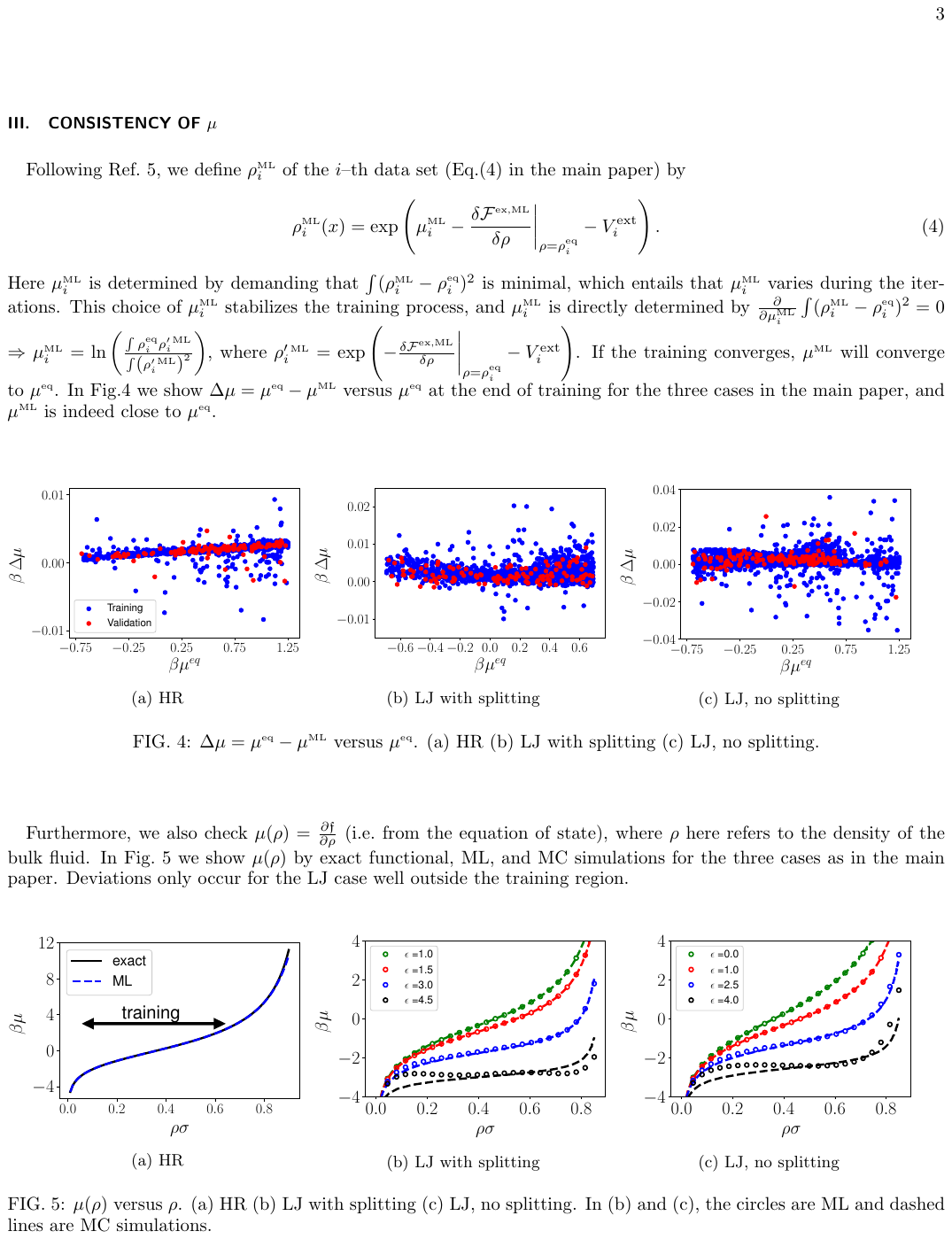}
\end{figure}
\begin{figure}
\includegraphics[page=4,width=\textwidth]{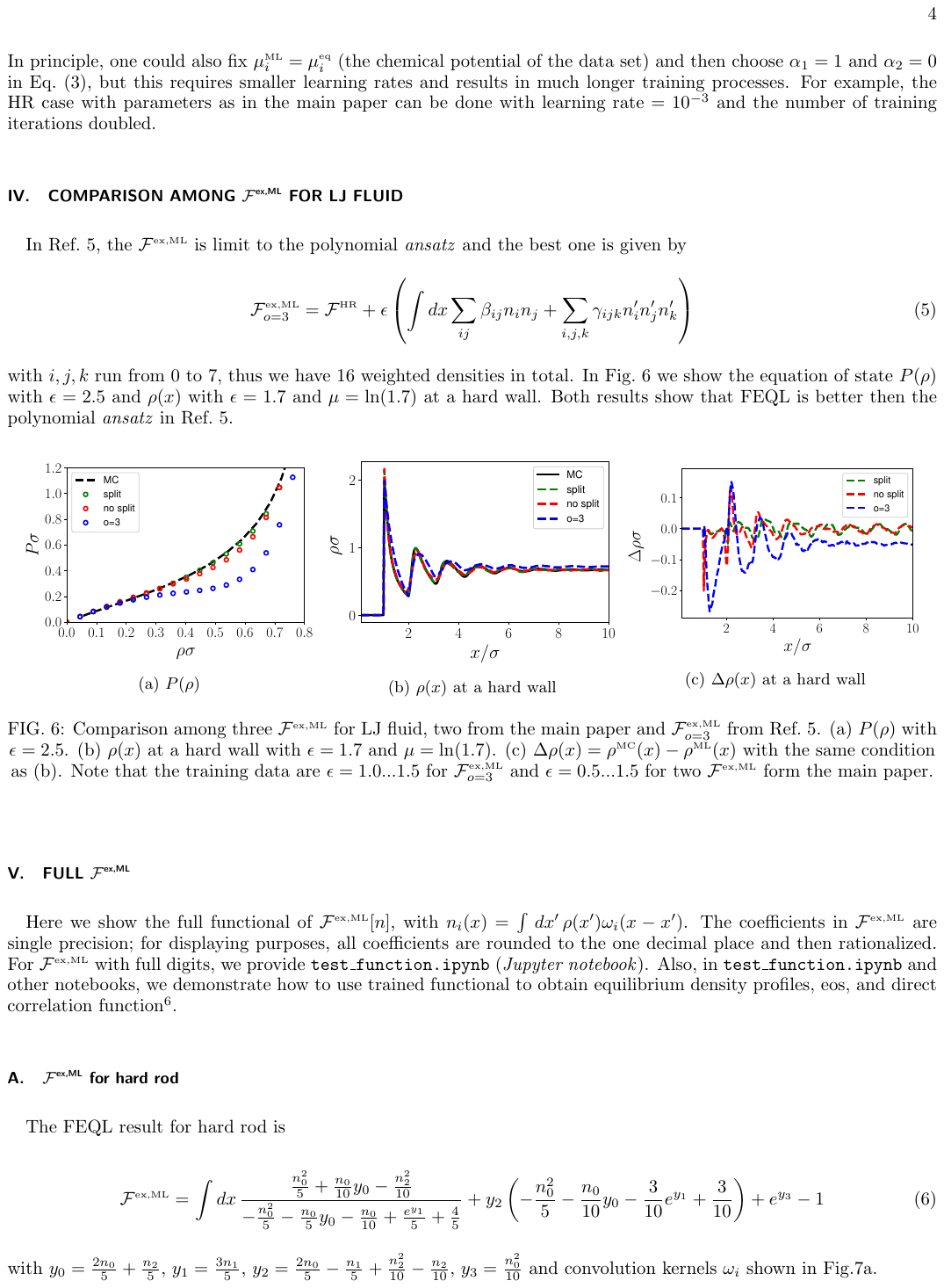}
\end{figure}
\begin{figure}
\includegraphics[page=5,width=\textwidth]{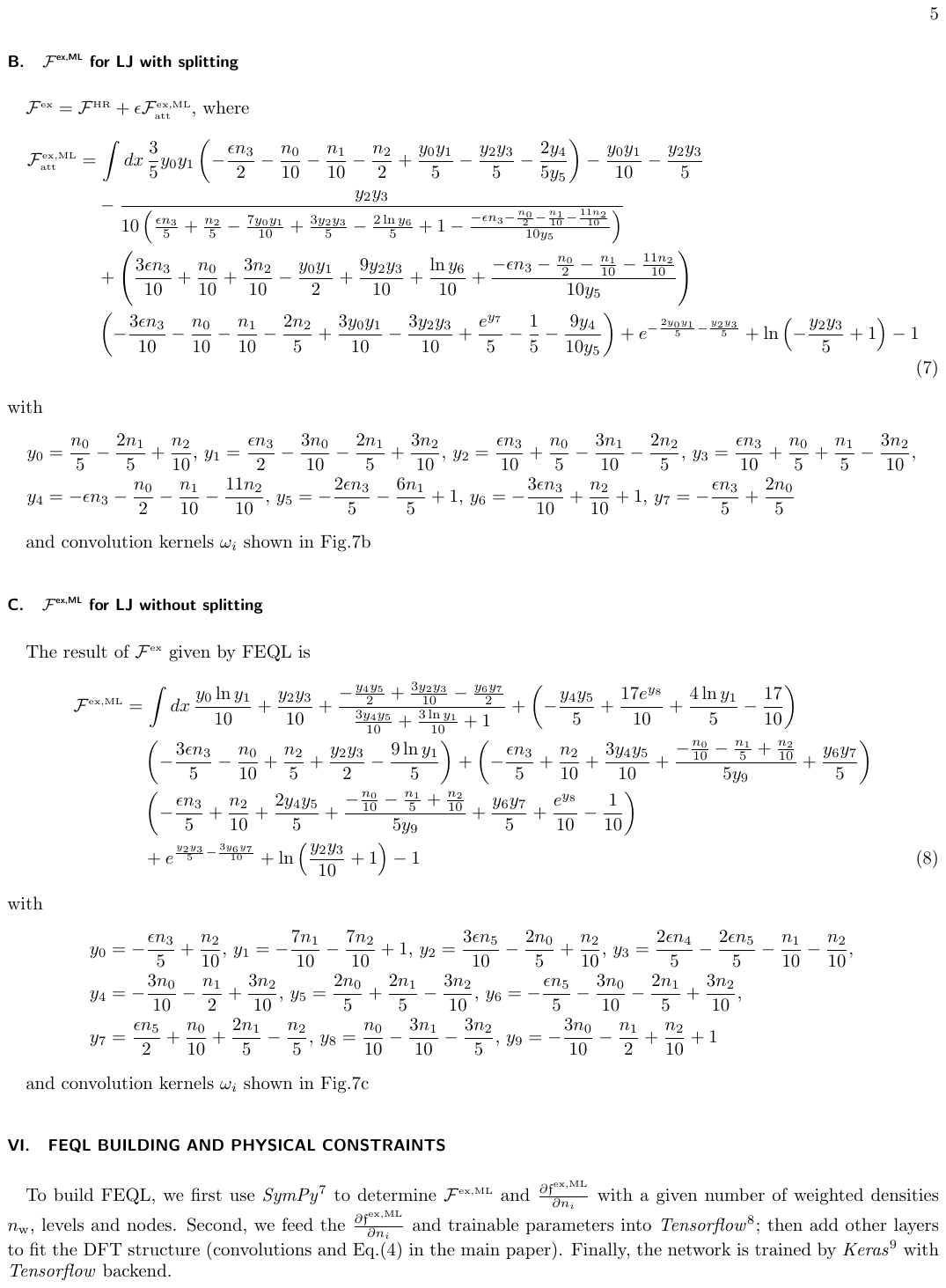}
\end{figure}
\begin{figure}
\includegraphics[page=6,width=\textwidth]{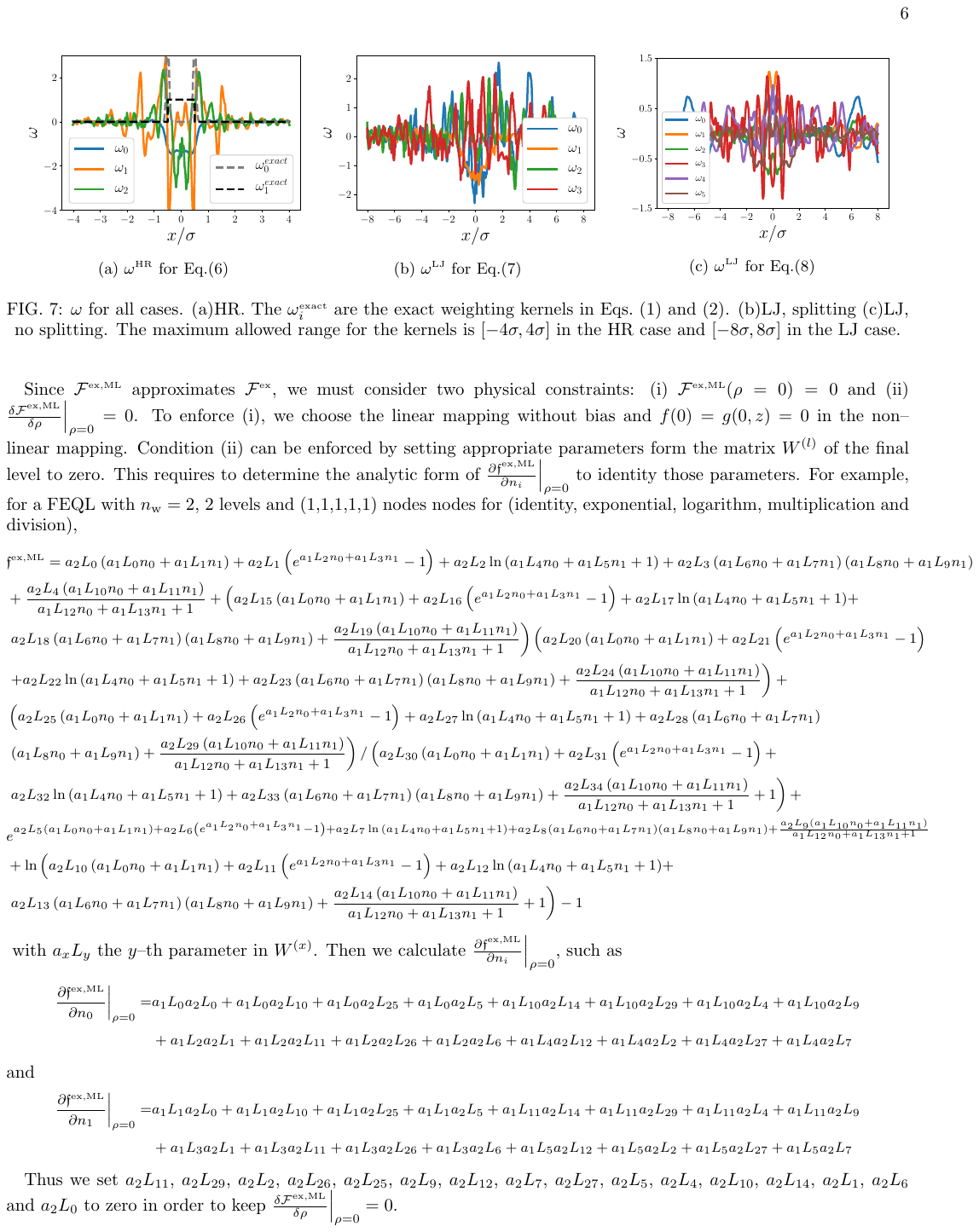}
\end{figure}
\begin{figure}
\includegraphics[page=7,width=\textwidth]{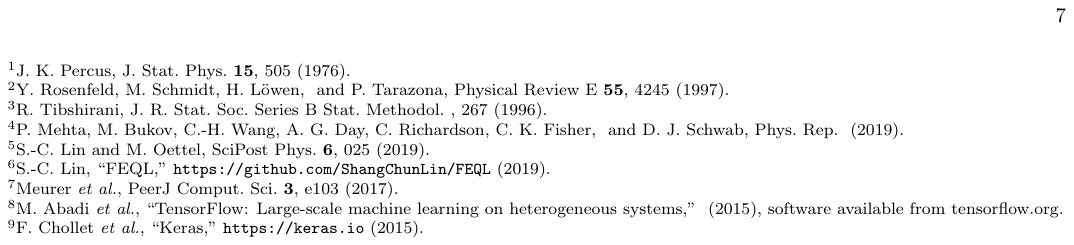}
\end{figure}

\end{document}